\documentclass[%
superscriptaddress,
preprint,
 amsmath,amssymb,
 aps,
prl,
longbibliography
]{revtex4-1}

\usepackage{xcolor}

\usepackage{braket}
\usepackage{graphicx}
\usepackage{dcolumn}
\usepackage{bm}
\usepackage{soul}

\begin{document}

\title{Atomic-scale control of spin transport in exchange-coupled molecules}

\author{A. F\'{e}tida}
\affiliation{Universit\'{e} de Strasbourg, CNRS, IPCMS, UMR 7504, F-67000 Strasbourg, France\looseness=-1}
\author{R. Robles}
\affiliation{Centro de Física de Materiales CFM/MPC (CSIC-UPV/EHU), Paseo Manuel de Lardizabal 5, 20018 Donostia-San Sebasti\'an, Spain\looseness=-1}
\author{F. Scheurer}
\author{M. Romeo}
\affiliation{Universit\'{e} de Strasbourg, CNRS, IPCMS, UMR 7504, F-67000 Strasbourg, France\looseness=-1}
\author{J. Kröger}
\affiliation{Institut für Physik, Technische Universität Ilmenau, D-98693 Ilmenau, Germany}
\author{N. Lorente}
\affiliation{Centro de Física de Materiales CFM/MPC (CSIC-UPV/EHU), Paseo Manuel de Lardizabal 5, 20018 Donostia-San Sebasti\'an, Spain\looseness=-1}
\affiliation{Donostia International Physics Center (DIPC), 20018 Donostia-San Sebasti\'{a}n, Spain\looseness=-1}
\author{L. Limot}
\email{limot@ipcms.unistra.fr}
\affiliation{Universit\'{e} de Strasbourg, CNRS, IPCMS, UMR 7504, F-67000 Strasbourg, France\looseness=-1}

\begin{abstract}

Spin transport between coupled spins is governed by tunneling and exchange interactions, yet their microscopic relation in real space has remained experimentally elusive. To address this, we functionalize the tip of a scanning tunneling microscope with a single molecule and use it as a movable spin sensor positioned above a second molecule on a ferromagnetic surface, forming an exchange-coupled molecular junction. From the same spectroscopic dataset, we independently extract the junction conductance, exchange coupling, and spin polarization with atomic-scale resolution. Conductance and exchange display nearly identical spatial dependences, consistent with a minimal Hubbard description in which both are governed by the same hopping matrix between the molecular frontier orbitals. In contrast, the spin polarization depends on the orbital composition and the relative weight of the resulting transport channels. These results establish a real-space connection between tunneling, exchange, and spin transport in magnetically coupled molecules.

\end{abstract}
\date{\today}


 \maketitle

\section{INTRODUCTION}

Transport between coupled quantum spins is governed by tunneling, which mediates spin transfer, and by exchange interactions, which couple their magnetic moments. Together, these processes are fundamental to spin manipulation at the nanoscale. In semiconductor quantum dots, for instance, tunnel coupling controls exchange interactions and enables two-qubit gates~\cite{Petta2005,Veldhorst2015}. In molecular spintronics, spin-polarized currents flow through organic bridges, while magnetic coupling can be engineered through chemical design~\cite{Barraud2010,Sanvito2010}. Atomic spin structures investigated by scanning tunneling microscopy (STM) with a spin-polarized tip have revealed how exchange interactions govern spin excitations and magnetic dynamics at the atomic scale~\cite{Loth2012,Yan2014}. However, directly probing how tunneling, exchange, and spin polarization evolve across a molecular junction with atomic-scale spatial resolution remains experimentally challenging.


Exchange coupling across vacuum provides a route to elucidate this link at the atomic scale within a scanning probe microscope. In this approach, a spin-sensitive object attached to the tip apex is exchange-coupled to a surface spin while spectroscopic signatures are detected in the tunneling current. Atomic-scale exchange sensing has been demonstrated using magnetic exchange force microscopy~\cite{Kaiser2007,Hauptmann2020,Adachi2025}, Kondo spectroscopy~\cite{Bork2011,Choi2016,Garnier2020,Ternes2020}, spin-excitation spectroscopy~\cite{Yan2014,Czap2019,Verlhac2019,Aguirre2024}, and electron spin resonance detected in the tunneling current~\cite{Yang2019,Willke2019,Kovarik2024,Esat2024}. Scanning the sensing spin laterally across the surface enables real-space mapping of the exchange interaction across the vacuum junction. This strategy has revealed magnetic structures with atomic resolution, ranging from single atoms~\cite{Willke2019,Czap2025} to magnetic surfaces~\cite{Kaiser2007,Hauptmann2020,Adachi2025,Fetida2024,Fetida2025}. Extending such spin-sensitive measurements to coupled molecular systems is especially appealing because molecules provide discrete spin states and well-defined frontier orbitals, offering access to interactions with orbital specificity.


Here, we use nickelocene [Ni(C$_5$H$_5$)$_2$, hereafter Nc] to realize such a molecular junction. One molecule is attached to the apex of an STM tip~\cite{Ormaza2017a,Ormaza2017b,Kogler2024}, forming a quantum spin probe (with $S=1$) whose spin excitations provide a direct spectroscopic measure of the local magnetic interaction with the sample~\cite{Czap2019,Verlhac2019}. A second Nc molecule is adsorbed on a magnetic cobalt nanoisland grown on Cu(111). By laterally positioning the Nc-terminated tip above the surface-supported Nc molecule, the overlap between the two molecules can be controlled with submolecular precision, forming a tunable exchange-coupled molecular junction. Previous Nc–Nc studies have mainly focused on point spectroscopy~\cite{Czap2019,Waeckerlin2022,Song2024,Bae2025,Lorenz2026} or on conductance imaging, where electronic and magnetic contributions remain intertwined~\cite{Czap2019,Waeckerlin2022}. We show that the spectroscopic response of the Nc–Nc junction can instead be exploited to simultaneously extract, with submolecular spatial resolution, three observables: conductance through elastic tunneling, and exchange interaction and spin polarization through inelastic spin excitations. Conductance and exchange exhibit a nearly identical spatial dependence, consistent with a minimal two-orbital Hubbard description in which both are governed by the total hopping strength between the molecular frontier states. By contrast, the spin polarization reflects the orbital composition and relative weight of the resulting transport channels, which can be continuously tuned by lateral displacement of the Nc-tip sensor. 


\begin{figure}[t]
 \includegraphics[width=0.9\columnwidth]{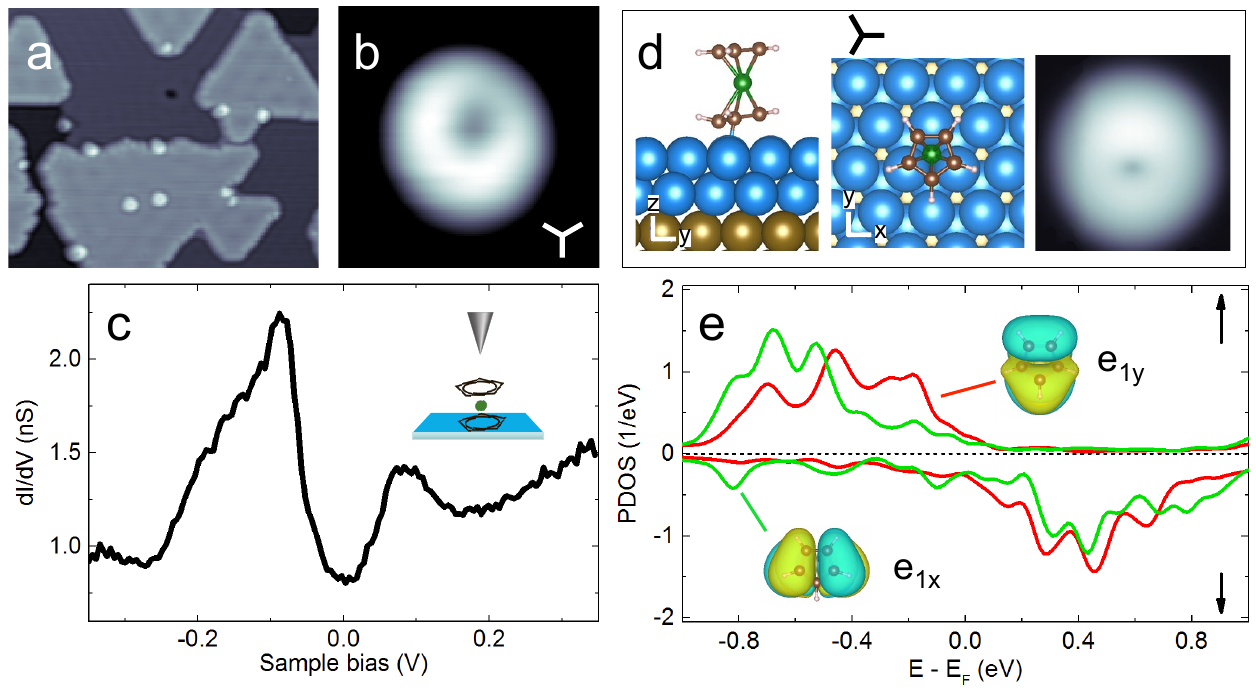}
 \caption{
(a) Co islands grown on Cu(111). Islands are two (2 ML) or, occasionally, one (1 ML) atomic layers high; bright protrusions correspond to Nc molecules. Image size: $9\times 11$~nm$^2$, bias: $50$~mV, current: $40$~pA. (b) Close-up image of Nc adsorbed on Co ($1.2\times1.2$~nm$^2$, $50$~mV, $40$~pA). The white axes indicate the three close-packed directions of the Co surface, determined from atomically resolved images~\cite{Verlhac2019}. (c) Constant-current $dI/dV$ spectrum acquired above the Cp ring of Nc, revealing molecular resonances. Feedback loop opened at $50$~mV, $50$~pA. Inset: schematic of the tunnel junction. (d) Relaxed adsorption geometry of Nc on a Co bilayer supported by Cu(111), obtained from DFT calculations (top and side views). The molecule adopts a tilted configuration. Right: corresponding Tersoff–Hamann simulated STM image. (e) Projected density of states (PDOS) for Nc/Co in the relaxed configuration shown in (d). Projections onto the $e_{1x}$ (green) and $e_{1y}$ (red) frontier molecular orbitals reveal a lifting of their degeneracy. Top and bottom panels: spin-up and spin-down PDOS. Insets: $e_{1x}$ and $e_{1y}$ orbitals.
}
\label{F1}
\end{figure}


\section{RESULTS AND DISCUSSION}

To build the Nc–Nc junction, we first characterize the adsorption geometry and electronic structure of Nc on cobalt (Nc/Co). Figure~\ref{F1}a shows a representative STM image of cobalt islands after exposure to Nc, acquired with a metallic W tip. The molecules adsorb both on the Co islands and along their step edges (additional molecules can also be positioned on the islands by tip-assisted manipulation; see section~S1 of the Supplementary Information). The ring-like contrast observed on the islands (Fig.~\ref{F1}b), together with an apparent height of $350 \pm 10$ pm, is characteristic of metallocenes~\cite{Bachellier2016,Garnier2020}, indicating adsorption with one cyclopentadienyl (C$_5$H$_5$, Cp) ring bound to the surface and the other exposed to vacuum. The exposed Cp ring exhibits a pronounced lateral asymmetry, appearing brighter on one side. This asymmetry is strongest at low bias (section~S2, Supplementary Information) and is mirror-symmetric with respect to an axis perpendicular to the close-packed rows of Co atoms. Figure~\ref{F1}c presents a typical constant-current $dI/dV$ spectrum acquired above the Cp ring. Compared to bare Co (section~S2, Supplementary Information), the surface $d$ states are replaced by molecular resonances, with prominent peaks near $-80$ mV and $+70$ mV, which we assigne to the $e_1$ frontier orbital of Nc (see below).

To complement the experimental data, we performed DFT calculations on a fully relaxed system consisting of an isolated Nc molecule adsorbed on a Co bilayer supported by a Cu slab (Fig.~\ref{F1}d; section~S1, Supplementary Information). The calculations show that Nc adsorbs in an upright configuration, with the lower Cp ring centered on an \textit{fcc} hollow site of the Co surface. The lowest carbon atom lies $193$ pm above the surface. One carbon atom of the Cp ring occupies a neighboring \textit{hcp} hollow site, inducing a tilt of the molecular axis by $4^\circ$ with respect to the surface normal. The partially occupied $e_1$ frontier orbitals of Nc, labeled $e_{1x}$ and $e_{1y}$, have their degeneracy partially lifted upon adsorption on Co, resulting in distinct PDOS for the two orbitals (Fig.~\ref{F1}e). Within the experimentally relevant energy window around the Fermi level, the tunneling current is dominated by the $e_{1y}$ orbital, giving rise to the asymmetric tilted ring-like contrast observed in the low-bias simulated Tersoff–Hamann image (Fig.~\ref{F1}d).

 \begin{figure}[t]
 \includegraphics[width=0.8\columnwidth]{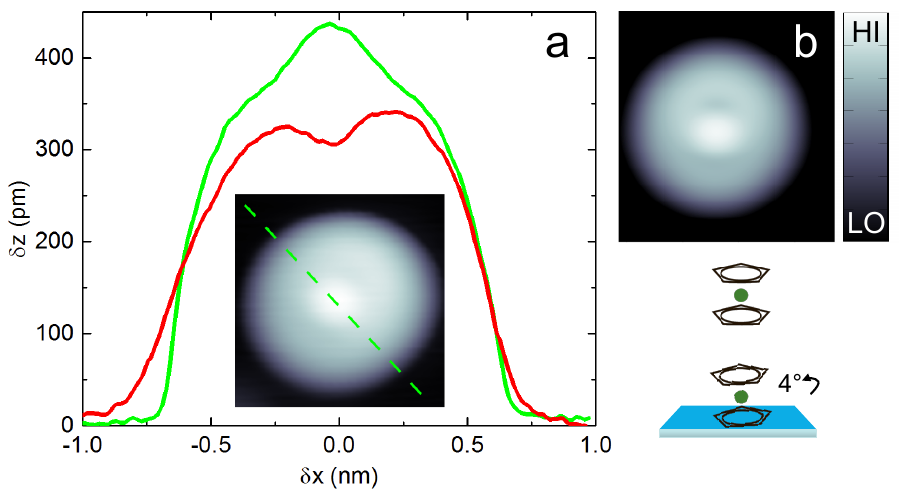}
  \caption{(a) Constant-current STM image and apparent-height profile of Nc/Co (2 ML) acquired with an Nc-tip ($50$ mV, $50$ pA, $1.5\times1.5$ nm$^2$). The solid green trace is the line profile taken along the dashed line in the image. For comparison, the apparent-height profile acquired with a metallic tip is shown as a solid red trace. (b) Constant-current images simulated within Bardeen’s tunneling formalism using the relaxed structure of Nc/Co (Fig.~\ref{F1}d;  (schematic on the bottom). Size: $2.1 \times 2.2~\mathrm{nm}^2$), bias: 0.22 V.
  }
\label{F2}
\end{figure} 



In the next step, we examine how coupling between the two Nc molecules reshapes the tunneling conductance channels. As shown in Fig.~\ref{F2}a, this is reflected in a spatial dependence of the conductance that differs markedly from that obtained with a metallic tip. Nc/Co appears nearly circular, with a pronounced central protrusion reaching $435 \pm 10$ pm, while its lateral extent remains comparable to that measured with a metal tip. A halo of weaker intensity surrounds the central maximum. Depending on the Nc-tip used, the position of the intensity maximum can exhibit lateral shifts (section~S3, Supplementary Information). To describe this conductance, we employ Bardeen’s tunneling formalism~\cite{Bardeen1961}, which explicitly incorporates the electronic structures of both tip and sample and thus goes beyond the Tersoff–Hamann approximation (section~S1, Supplementary Information). Within this framework, the tunneling current is determined by the overlap of tip and sample wavefunctions evaluated on a surface inside the vacuum barrier. The DFT-based simulation shown in Fig.~\ref{F2}b is computed using the relaxed geometry of Nc/Co(2 ML) described above. The simulated image reproduces the central intensity maximum as seen experimentally. The origin of the central maximum can be understood using Chen's derivative rules (Section~S4, Supplementary Information)~\cite{Chen1990}. In the tunneling regime, the $e_{1x}$ and $e_{1y}$ frontier orbitals of Nc decay into the vacuum as effective $d_{xz}$- and $d_{yz}$-like tunneling channels. Since the sample Nc molecule exhibits the same orbital symmetry, the tunneling matrix elements are determined by the overlap of the corresponding effective $d_{xz}$ and $d_{yz}$ channels across the junction. This overlap is maximal when the centers of the two molecules coincide, producing the pronounced central conductance maximum observed experimentally and reproduced by the simulations. Imposing a tilt angle (Section~S4, Supplementary Information) results in a laterally displaced maximum, hence the position of the intensity maximum reflects the tilt angle of the Nc molecule at the tip apex. In the following, we restrict the analysis to Nc tips with minimal tilt, without loss of generality.



 \begin{figure}[t]
 \includegraphics[width=0.85\columnwidth]{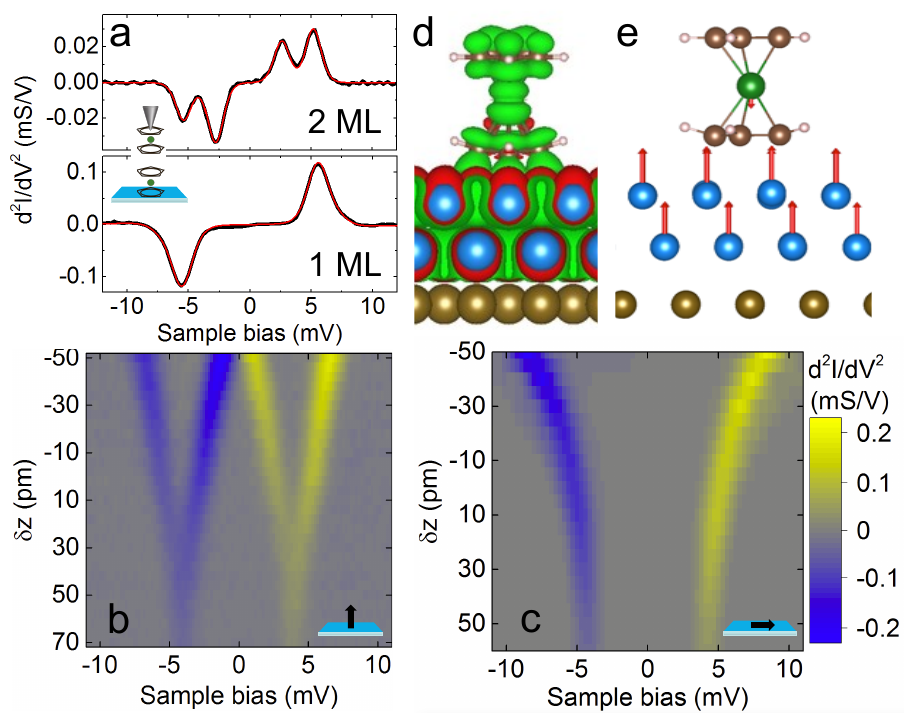}
  \caption{(a) Spin-excitation spectra of Nc/Co adsorbed on Co islands of two different thicknesses: 2 ML (top, black line) and 1 ML (bottom, black line). The feedback loop was opened at $12$ mV and $0.7$ nA for the 2 ML island, and at $12$ mV and $4$ nA for the 1 ML island. Red lines are simulations based on a dynamical scattering model~\cite{Ternes2015}. Inset: schematic representation of the tunnel junction used for the spin-sensitive measurements. Panels (b) and (c) present two-dimensional intensity plots showing a series of distance-dependent $d^2I/dV^2$ spectra acquired above Nc/Co on a 2 ML island (b) and a 1 ML island (c). A conductance of $G=5.2\cdot10^{-3}\,G_0$ (where $G_0=2e^2/h$ is the quantum of conductance), corresponds to $\delta z=0$. (d) Side view of an isosurface plot of the spin density for the DFT-optimized structure of Nc adsorbed on Co(2 ML). Red and green indicate spin-up and spin-down densities, respectively. (e) Calculated magnetization of the Nc molecule and the underlying Co atoms. Arrows indicate the orientation of the local magnetic moments.} 
\label{F3}
\end{figure}


Finally, before addressing the Nc–Nc junction specifically, we determine the magnetic properties of Nc/Co by comparing Nc adsorbed on Co bilayer and monolayer islands. The bilayer islands exhibit out-of-plane magnetization, whereas the monolayer islands are magnetized in plane~\cite{Fetida2024}. Figure~\ref{F3}a shows representative spin-excitation spectra acquired with the Nc-tip positioned above the center of Nc/Co on the two island types. In both cases, symmetric conductance features are observed around zero bias, characteristic of inelastic spin excitations of the Nc-tip. For Nc on the bilayer island, the excitation is split into two components, whereas for Nc on the monolayer island only a single feature is resolved.

The distance dependence of the spectra is shown in Figs.~\ref{F3}b,c (decreasing the tip displacement $\delta z$ corresponds to moving the tip closer to the surface). On the bilayer, reducing the tip–sample separation produces a continuous increase of the splitting between the two excitation features. On the monolayer, the same approach primarily shifts the excitation energy, without generating a resolved splitting. To rationalize these observations, we describe the Nc-tip by the effective spin Hamiltonian
\begin{equation}
\mathcal{H} = D S_z^2 - g \mu_\mathrm{B} \mathbf{B}_{\mathrm{ex}} \cdot \mathbf{S},
\label{eq1}
\end{equation}
where $D=4.0\pm0.1$ meV is the intrinsic easy-plane magnetic anisotropy of Nc on the tip, $g=1.89$~\cite{Czap2019}, and $\mathbf{B}_{\mathrm{ex}}$ is the exchange field exerted by Nc/Co on the spin of the Nc-tip. Within this model, a field component along the anisotropy axis splits the excitation, whereas a transverse component mainly shifts its energy. The contrasting spectra on bilayer and monolayer islands are therefore consistently explained if $\mathbf{B}_{\mathrm{ex}}$ follows the magnetization direction of the underlying Co island: out of plane for the bilayer and in plane for the monolayer. The magnitude of $B_{\mathrm{ex}}$ increases exponentially as the tip–sample separation is reduced (section~S5, Supplementary Information). Nc/Co has thus a magnetization locked to that of the underlying island through exchange coupling, while still acting as a localized molecular spin in the junction. We note that no spin excitation of surface-adsorbed Nc is resolved experimentally (Fig.~\ref{F1}c), likely because of hybridization with Co. This situation differs from previously reported Nc–Nc junctions~\cite{Ormaza2017a,Czap2019,Song2024,Bae2025}, where both molecules showed detectable spin excitations in the same energy window. Here, the junction effectively contains a single spin-active probe, the Nc-tip, as captured by Eq.~\ref{eq1}.


To corroborate these findings, we performed spin-polarized DFT calculations using the relaxed adsorption geometry of Nc on Co(2 ML). The total magnetic moment of Nc amounts to $1.28~\mu_\mathrm{B}$ ($0.77~\mu_\mathrm{B}$ on the Ni atom), reduced compared to the $2~\mu_\mathrm{B}$ of the free-standing molecule, indicating weak chemisorption. The spin-density isosurface (Fig.~\ref{F3}d) reveals an antiferromagnetic alignment between the Nc molecule and the Co substrate. Magnetocrystalline anisotropy calculations identify the surface normal as the easy axis, with the out-of-plane configuration favored by $17.7$~meV (Fig.~\ref{F3}e; section~S1, Supplementary Information). 
The calculations show a strong antiparallel alignment of Nc and Co spins. The parallel alignment is typically unstable in the calculation and it reverts to an antiparallel configuration or to a low spin case (section~S1, Supplementary Information). This shows that the molecular magnetization of Nc is locked antiparallel to that of the Co island by exchange coupling, in agreement with observations.

 \begin{figure}[t]
 \includegraphics[width=0.85\columnwidth]{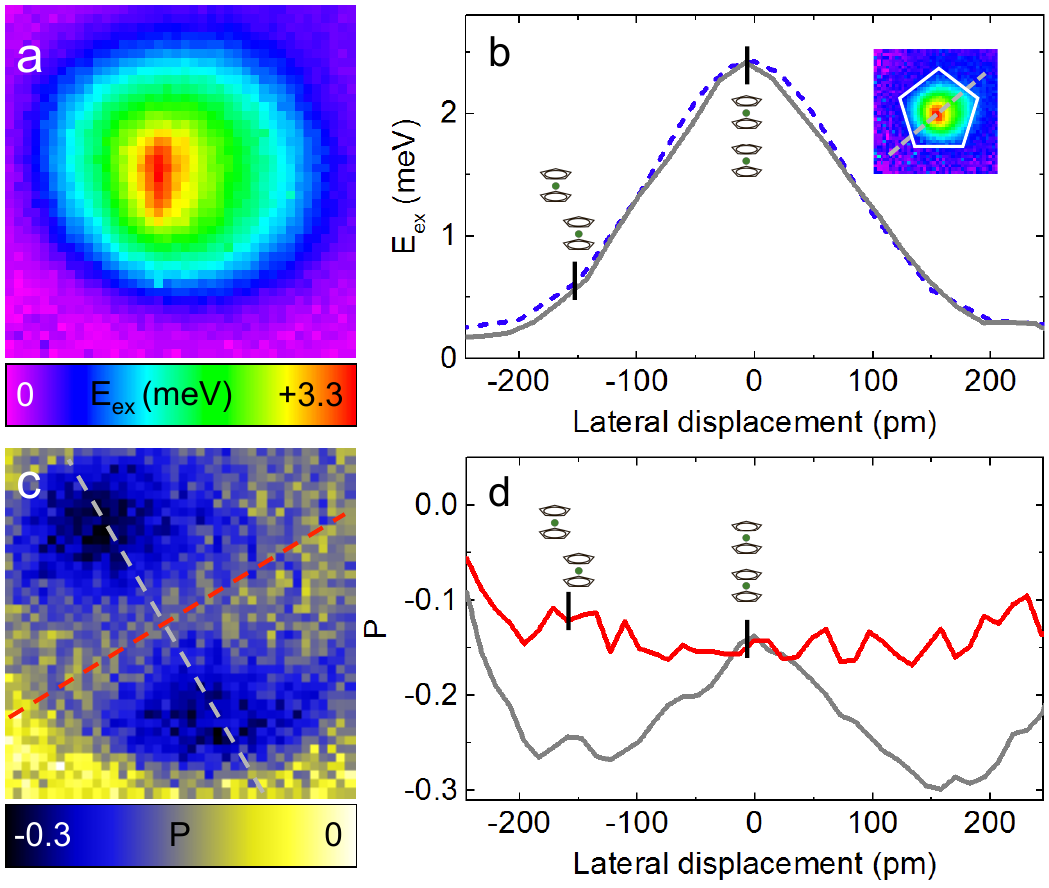}
  \caption{(a) Exchange-energy map of the Nc-Nc junction, measured over $0.51 \times 0.51~\mathrm{nm}^2$ at constant height after opening the feedback at $z=-210~\mathrm{pm}$. (b) Line profile of $E_\mathrm{ex}$ extracted along the dashed line in the inset (gray), compared with the DFT-computed profile (blue; see Supplementary Information). The DFT profile was scaled down to account for the higher absolute exchange energies arising from uncertainties in the intermolecular distance and the collinear spin approximation. Inset: $E_\mathrm{ex}$ map acquired with a different Nc tip over $0.65 \times 0.65~\mathrm{nm}^2$; the Cp-ring size is superimposed as a guide to the eye. (c) Spin-polarization map $P$ acquired simultaneously with (a). (d) Profiles of $P$ extracted along the red and green directions indicated in (c). The molecular sketches in panels (b) and (d) illustrate representative junction geometries at selected lateral displacements, drawn to scale with respect to the Cp-ring radius.} 
\label{F4}
\end{figure}

The lateral dependence of the Nc-Nc interaction was investigated by scanning the Nc tip above Nc/Co at constant tip-surface distance with the feedback loop disabled. The chosen distance ensured a non-repulsive Nc-Nc regime while maintaining junction stability (section~S6, Supplementary Information). At each pixel, the spin-excitation spectra were analyzed using a dynamical scattering model~\cite{Ternes2015}. The positions of the excitation thresholds provide the local exchange field and thus the exchange energy $E_\mathrm{ex}(x,y)=g\mu_B B_\mathrm{ex}(x,y)$, whereas the relative spin-excitation amplitudes at positive and negative bias yield the spin polarization $P(x,y)$ of the tunneling current.

Figure~\ref{F4} presents a typical exchange map, showing a maximum of
$3.3$~meV at the center of Nc/Co. A second map acquired with another tip shows a similar behavior (inset of Fig.~\ref{F4}b). The profile in Fig.~\ref{F4}b yields a FWHM of $190$~pm, significantly smaller than the apparent lateral extent of Nc/Co, yet close to the $220$~pm diameter of a Cp ring. This demonstrates that the exchange interaction is confined laterally to a region of molecular dimensions, comparable to a single Cp ring.


The need to explicitly account for the molecular structure of the Nc-tip in the conductance maps (Fig.~\ref{F2}) motivates a similar approach for the exchange images. We therefore model the junction using two vertically stacked gas-phase Nc molecules separated by $350$ pm, with the upper molecule laterally displaced relative to the lower one (section~S1, Supplementary Information). For each lateral position, the structure is relaxed and the exchange energy is calculated as $\Delta E^\mathrm{Nc-Nc}=E^\mathrm{Nc-Nc}_\mathrm{P}-E^\mathrm{Nc-Nc}_\mathrm{AP}$, where $\mathrm{P}$ ($\mathrm{AP}$) denotes parallel (antiparallel) alignment of the molecular spins. Despite neglecting both the Co substrate and the metallic tip, this minimal model captures the main experimental observations. The exchange interaction is antiferromagnetic, reaches its maximum magnitude for vertically aligned molecules, and decreases upon lateral displacement, yielding a spatial profile in agreement with experiment (Fig.~\ref{F4}b). 


Strikingly, the conductance and exchange maps, obtained independently from elastic and inelastic tunneling processes, display a nearly identical spatial dependence. This correspondence can be captured by a minimal two-site Hubbard description~\cite{Waeckerlin2022}. In the present case, transport occurs through the frontier $e_1$ states of
the two Nc molecules ($e_{1x}$ and $e_{1y}$). The coupling between the two $e_1$ manifolds is described by a hopping matrix $T(\mathbf R)$, whose elements depend on the lateral displacement $\mathbf R$ of the Nc-tip relative to Nc/Co. This matrix defines two effective transport channels with hopping amplitudes $t_1(\mathbf R)$ and $t_2(\mathbf R)$, corresponding to the principal coupling paths rather than individual $e_1$ orbitals (section~S7, Supplementary Information). In the tunneling regime, where the on-site Coulomb interaction satisfies $U\gg |t_i|$, the Hubbard model reduces to an effective Heisenberg interaction. The antiferromagnetic exchange obtained from DFT is consistent with this Hubbard superexchange mechanism in the large-$U$ limit, for which $E_\mathrm{ex}(\mathbf R)\propto (|t_1(\mathbf R)|^2+|t_2(\mathbf R)|^2)/U$. The exchange map therefore reflects the total hopping strength between the two molecular $e_1$ manifolds. The same transport channels determine elastic tunneling, yielding a conductance $G(\mathbf R)\propto\Gamma_1|t_1(\mathbf R)|^2+\Gamma_2|t_2(\mathbf R)|^2$, where $\Gamma_i$ accounts for the electronic spectral weight of channel $i$, including the hybridization of the molecular states with their respective electrodes. The experimentally observed similarity between $G(\mathbf R)$ and $E_\mathrm{ex}(\mathbf R)$ indicates that both conductance and exchange are governed predominantly by the same quantity $|t_1(\mathbf R)|^2+|t_2(\mathbf R)|^2$, with, consequently, comparable effective hybridizations $\Gamma_1\simeq\Gamma_2$. The exchange map can thus be viewed as the spin counterpart of the conductance map, with both originating from the same hopping matrix between the molecular frontier states. This situation differs from exchange imaging on metallic surfaces, where the Nc-tip probes a continuum of electronic states. In that case, tunneling involves multiple electronic channels, and lateral variations of the exchange interaction primarily reflect the spatial distribution of spin density within the magnetic surface~\cite{Fetida2024,Fetida2025}.

Finally, we investigate spin transport in the Nc–Nc junction at the Fermi level. The spin-transport information is extracted simultaneously with the exchange energy by recording the amplitudes of the spin-excitation features at each pixel of the image~\cite{Loth2010b,Fetida2024}, thereby yielding a spatial map of the spin polarization at the Fermi energy, shown in Fig.~\ref{F4}c. A pronounced lateral dependence is observed as the relative position of the two Nc molecules is varied, whereas no dependence on tip-sample distance is found (section~S5, Supplementary Information). The spin polarization $P$ exhibits a two-lobe feature, with some variation in contrast depending on the specific Nc-tip used (Supplementary Information). In the centered configuration, where the molecular axes are aligned, we obtain $P = -0.15$ (Fig.~\ref{F4}d). In off-center configurations, the polarization decreases to $P = -0.30$ across the two lobes, while remaining essentially unchanged along the orthogonal direction. This finite polarization contrasts with pristine Co bilayers ($P= 0$)~\cite{Verlhac2019,Fetida2024}, indicating the emergence of a molecular spinterface at the Nc/Co interface~\cite{Barraud2010,Atodirisei2010,Sanvito2010}.

The measured spin polarization can be rationalized by considering the spin-dependent character of the two transport channels derived from the $e_1$ molecular states. The junction polarization reads $P(\mathbf{R})=\left[\lambda(\mathbf{R})p_1+p_2\right]/\left[\lambda(\mathbf{R})+1\right]$ (section~S7, Supplementary Information), where $p_i$ is the spin polarization associated with channel $i$. The parameter $\lambda(\mathbf{R})=|t_1(\mathbf{R})|^2/|t_2(\mathbf{R})|^2$ represents the relative transmission weight of the two channels and is controlled by the position-dependent hopping matrix between the two molecules. For a centered junction, the molecular symmetry results in comparable channel contributions, $\lambda\simeq1$, and the measured spin polarization corresponds approximately to the average of the two channel polarizations. Upon lateral displacement, changes in the hopping matrix modify both the relative weights and the orbital composition of the two transport channels, causing the measured polarization to evolve towards the channel contributing most strongly to transport. The two-lobe feature observed experimentally therefore indicates a strong difference in spin polarization between the two $e_1$-derived channels, as also suggested by the spin-PDOS of Nc/Co (Fig.~\ref{F2}e), and is consistent with one weakly polarized channel and another carrying a negative polarization of about $-0.3$. The lateral tip displacement therefore does not simply select a molecular orbital, but continuously modifies the composition and relative weight of spin-polarized transport channels within the molecular junction.

\section{CONCLUSIONS}
We have imaged spin transport in a junction formed by two exchange-coupled nickelocene molecules using a single-molecule quantum sensor. By simultaneously mapping conductance, exchange, and spin polarization from the same spectroscopic dataset, we establish a direct real-space link between elastic tunneling and correlated spin interactions within a single molecular junction. This simultaneous access is essential for disentangling the distinct roles of orbital coupling and spin-dependent transport channels at the atomic scale. Conductance and exchange are governed by the same hopping matrix between the molecular frontier states, whereas the spin polarization reflects the relative contribution and spin character of the resulting transport channels. These findings demonstrate that spin transport across magnetically coupled molecules can be controlled through the orbital structure of the molecular junction, providing a framework for designing spin functionality at the molecular scale.\\


\section{ACKNOWLEDGEMENTS}
The Strasbourg authors gratefully acknowledge V.~Speisser for technical support and maintenance of the experimental setup. They also acknowledge support from the EU’s Horizon 2020 research and innovation programme under the Marie Skłodowska-Curie grant 847471 and from project ANR-23-CE09-0036 funded by the ANR. This work is supported by France 2030 government investment plan managed by the French National Research Agency under grant reference PEPR SPIN–SPINMAT ANR-22-EXSP-0007. R.R. and N.L. are grateful for financial support from projects PID2021-127917NB-I00 funded by MCIN/AEI/ 10.13039/501100011033, from project IT-1527-22 funded by the Basque Government, and from project ESiM 101046364 funded by the European Union.

\clearpage
\bibliography{references}

@article{Petta2005,
    author = {J. R. Petta  and A. C. Johnson  and J. M. Taylor  and E. A. Laird  and A. Yacoby  and M. D. Lukin  and C. M. Marcus  and M. P. Hanson  and A. C. Gossard },
    title = {Coherent Manipulation of Coupled Electron Spins in Semiconductor Quantum Dots},
    journal = {Science},
    volume = {309},
    number = {5744},
    pages = {2180-2184},
    year = {2005},
    doi = {10.1126/science.1116955},
    URL = {https://www.science.org/doi/abs/10.1126/science.1116955}
}

@article{Veldhorst2015,
	author = {Veldhorst, M. and Yang, C. H. and Hwang, J. C. C. and Huang, W. and Dehollain, J. P. and Muhonen, J. T. and Simmons, S. and Laucht, A. and Hudson, F. E. and Itoh, K. M. and Morello, A. and Dzurak, A. S.},
	date = {2015/10/01},
	doi = {10.1038/nature15263},
	id = {Veldhorst2015},
	isbn = {1476-4687},
	journal = {Nature},
	number = {7573},
	pages = {410--414},
	title = {A two-qubit logic gate in silicon},
	url = {https://doi.org/10.1038/nature15263},
	volume = {526},
	year = {2015}}

@article{Barraud2010,
	author = {Barraud, C. and Seneor, P. and Mattana, R. and Fusil, S. and Bouzehouane, K. and Deranlot, C. and Graziosi, P. and Hueso, L. and Bergenti, I. and Dediu, V. and Petroff, F. and Fert, A.},
	doi = {10.1038/nphys1688},
	journal = {Nat. Phys.},
	number = {8},
	pages = {615--620},
	title = {Unravelling the role of the interface for spin injection into organic semiconductors},
	url = {https://doi.org/10.1038/nphys1688},
	volume = {6},
	year = {2010}}

@article{Sanvito2010,
	author = {Sanvito, S.},
	journal = {Nat. Phys.},
	number = {8},
	pages = {562--564},
	title = {The rise of spinterface science},
	volume = {6},
	year = {2010}
}

@article{Loth2012,
    author = {Loth, S. and Baumann, S. and Lutz, C. P. and Eigler D.M. and Heinrich, A. J. },
    title = {Bistability in Atomic-Scale Antiferromagnets},
    journal = {Science},
    volume = {335},
    number = {6065},
    pages = {196-199},
    year = {2012},
    doi = {10.1126/science.1214131},
    URL = {https://www.science.org/doi/abs/10.1126/science.1214131}
}

@article{Yan2014,
	Author = {Yan, S. and Choi, D.-J. and Burgess, J. A. J. and Rolf-Pissarczyk, S. and Loth, S.},
	Journal = {Nat. Nanotechnol.},
	Month = {12},
	Pages = {40},
	Title = {Control of quantum magnets by atomic exchange bias},
	Volume = {10},
	Year = {2014}
	}

@article{Kaiser2007,
	author = {Kaiser, U. and Schwarz, A. and Wiesendanger, R.},
	date = {2007/03/01},
	doi = {10.1038/nature05617},
	id = {Kaiser2007},
	isbn = {1476-4687},
	journal = {Nature},
	number = {7135},
	pages = {522--525},
	title = {Magnetic exchange force microscopy with atomic resolution},
	url = {https://doi.org/10.1038/nature05617},
	volume = {446},
	year = {2007}}

@article{Hauptmann2020,
	author = {Hauptmann, N. and Haldar, S. and Hung, T.-C. and Jolie, W. and Gutzeit, M. and Wegner, D. and Heinze, S. and Khajetoorians, A. A.},
	date = {2020/03/05},
	doi = {10.1038/s41467-020-15024-2},
	id = {Hauptmann2020},
	isbn = {2041-1723},
	journal = {Nat. Commun.},
	number = {1},
	pages = {1197},
	title = {Quantifying exchange forces of a spin spiral on the atomic scale},
	url = {https://doi.org/10.1038/s41467-020-15024-2},
	volume = {11},
	year = {2020}}

@article{Adachi2025,
    author ="Adachi, Y. and Yasui, Y. and Iiyama, A. and Kurahashi, W. and Nagase, R. and Sugimoto, Y.",
    title  ="Probing the spin spiral in {F}e chains on {I}r(001) using magnetic exchange force microscopy",
    journal  ="Nanoscale Horiz.",
    year  ="2025",
    volume  ="10",
    issue  ="8",
    pages  ="1653-1659",
    publisher  ="The Royal Society of Chemistry",
    doi  ="10.1039/D5NH00162E",
    url  ="http://dx.doi.org/10.1039/D5NH00162E"
}

@article{Bork2011,
	Author = {Bork, J. and Zhang, Y.-h. and Diekh{\"o}ner, L. and Borda, L. and Simon, P. and Kroha, J. and Wahl, P. and Kern, K.},
	Date = {2011/08/28/online},
	Day = {28},
	Journal = {Nat. Phys.},
	L3 = {10.1038/nphys2076; https://www.nature.com/articles/nphys2076#supplementary-information},
	M3 = {Article},
	Month = {08},
	Pages = {901},
	Publisher = {Nature Publishing Group SN  -},
	Title = {A tunable two-impurity Kondo system in an atomic point contact},
	Ty = {JOUR},
	Url = {http://dx.doi.org/10.1038/nphys2076},
	Volume = {7},
	Year = {2011}
	}

@article{Choi2016,
	author = {Choi, D.-J. and Guissart, S. and Ormaza, M. and Bachellier, N. and Bengone, O. and Simon, P. and Limot, L.},
	title = {Kondo Resonance of a {C}o Atom Exchange Coupled to a Ferromagnetic Tip},
	journal = {Nano Lett.},
	volume = {16},
	number = {10},
	pages = {6298-6302},
	year = {2016},
	doi = {10.1021/acs.nanolett.6b02617},
	URL = {http://dx.doi.org/10.1021/acs.nanolett.6b02617}
	}

@article{Garnier2020,
	author = {Garnier, L. and Verlhac, B. and Abufager, P. and Lorente, N. and Ormaza, M. and Limot, L.},
	date = {2020/11/11},
	doi = {10.1021/acs.nanolett.0c03271},
	isbn = {1530-6984},
	journal = {Nano Lett.},
	month = {11},
	number = {11},
	pages = {8193--8199},
	publisher = {American Chemical Society},
	title = {The {K}ondo Effect of a Molecular Tip As a Magnetic Sensor},
	type = {doi: 10.1021/acs.nanolett.0c03271},
	url = {https://doi.org/10.1021/acs.nanolett.0c03271},
	volume = {20},
	year = {2020},
	year1 = {2020}}

@article{Ternes2020,
    title = {Sensing the Spin of an Individual {C}e Adatom},
    author = {Ternes, M. and Lutz, C. P. and Heinrich, A. J. and Schneider, W.-D.},
    journal = {Phys. Rev. Lett.},
    volume = {124},
    issue = {16},
    pages = {167202},
    numpages = {6},
    year = {2020},
    month = {Apr},
    publisher = {American Physical Society},
    doi = {10.1103/PhysRevLett.124.167202},
    url = {https://link.aps.org/doi/10.1103/PhysRevLett.124.167202}
}

@article {Czap2019,
	author = {Czap, G. and Wagner, P. J. and Xue, F. and Gu, L. and Li, J. and Yao, J. and Wu, R. and Ho, W.},
	title = {Probing and imaging spin interactions with a magnetic single-molecule sensor},
	volume = {364},
	number = {6441},
	pages = {670--673},
	year = {2019},
	doi = {10.1126/science.aaw7505},
	publisher = {American Association for the Advancement of Science},
	issn = {0036-8075},
	URL = {https://science.sciencemag.org/content/364/6441/670},
	journal = {Science}
}

@article {Verlhac2019,
	author = {Verlhac, B. and Bachellier, N. and Garnier, L. and Ormaza, M. and Abufager, P. and Robles, R. and Bocquet, M.-L. and Ternes, M. and Lorente, N. and Limot, L.},
	title = {Atomic-scale spin sensing with a single molecule at the apex of a scanning tunneling microscope},
	volume = {366},
	number = {6465},
	pages = {623--627},
	year = {2019},
	doi = {10.1126/science.aax8222},
	publisher = {American Association for the Advancement of Science},
	issn = {0036-8075},
	URL = {https://science.sciencemag.org/content/366/6465/623},
	journal = {Science}
}

@article{Aguirre2024,
    author = {Aguirre, A. and Pinar Solé, A. and Soler Polo, D. and González-Orellana, C. and Thakur, A. and Ortuzar, J. and Stesovych, O. and Kumar, M. and Peña-Díaz, M. and Weber, A. and Tallarida, M. and Dai, J. and Dreiser, J. and Muntwiler, M. and Rogero, C. and Pascual, J. I. and  Jelínek, P. and Ilyn, M. and Corso, M.},
    title = {Ferromagnetic Order in {2D} Layers of Transition Metal Dichlorides},
    journal = {Adv. Mater.},
    volume = {36},
    number = {28},
    pages = {2402723},
    doi = {https://doi.org/10.1002/adma.202402723},
    url = {https://advanced.onlinelibrary.wiley.com/doi/abs/10.1002/adma.202402723},
    year = {2024}
}

@article{Yang2019,
    title = {Tuning the Exchange Bias on a Single Atom from 1 {mT} to 10 {T}},
    author = {Yang, K. and Paul, W. and Natterer, F. D. and Lado, J. L. and Bae, Y. and Willke, P. and Choi, T. and Ferr\'on, A. and Fern\'andez-Rossier, J. and Heinrich, A. J. and Lutz, C. P.},
    journal = {Phys. Rev. Lett.},
    volume = {122},
    issue = {22},
    pages = {227203},
    numpages = {6},
    year = {2019},
    month = {Jun},
    publisher = {American Physical Society},
    doi = {10.1103/PhysRevLett.122.227203},
    url = {https://link.aps.org/doi/10.1103/PhysRevLett.122.227203}
}

@article{Willke2019,
	author = {Willke, P. and Yang, K. and Bae, Y. and Heinrich, A. J. and Lutz, C. P.},
	date = {2019/10/01},
	doi = {10.1038/s41567-019-0573-x},
	id = {Willke2019},
	isbn = {1745-2481},
	journal = {Nat. Phys.},
	number = {10},
	pages = {1005--1010},
	title = {Magnetic resonance imaging of single atoms on a surface},
	url = {https://doi.org/10.1038/s41567-019-0573-x},
	volume = {15},
	year = {2019}}

@article{Kovarik2024,
    author = {Kovarik, S. and Schlitz, R. and Vishwakarma, A. and Ruckert, D. and Gambardella, P. and Stepanow, S.},
    title = {Spin torque-driven electron paramagnetic resonance of a single spin in a pentacene molecule},
    journal = {Science},
    volume = {384},
    number = {6702},
    pages = {1368-1373},
    year = {2024},
    doi = {10.1126/science.adh4753},
    URL = {https://www.science.org/doi/abs/10.1126/science.adh4753}
}

@article{Esat2024,
	author = {Esat, T. and Borodin, D. and Oh, J. and Heinrich, A. J. and Tautz, F. S. and Bae, Y. and Temirov, R.},
	date = {2024/10/01},
	doi = {10.1038/s41565-024-01724-z},
	id = {Esat2024},
	isbn = {1748-3395},
	journal = {Nat. Nanotechnol.},
	number = {10},
	pages = {1466--1471},
	title = {A quantum sensor for atomic-scale electric and magnetic fields},
	url = {https://doi.org/10.1038/s41565-024-01724-z},
	volume = {19},
	year = {2024}}

@article{Czap2025,
	author = {Czap, G. and Noh, K. and Velasco, J. Jr. and Macfarlane, R. M. and Brune, H. and Lutz, C. P.},
	date = {2025/01/28},
	doi = {10.1021/acsnano.4c14327},
	isbn = {1936-0851},
	journal = {ACS Nano},
	journal1 = {ACS Nano},
	journal2 = {ACS Nano},
	month = {01},
	number = {3},
	pages = {3705--3713},
	publisher = {American Chemical Society},
	title = {Direct Electrical Access to the Spin Manifolds of Individual Lanthanide Atoms},
	type = {doi: 10.1021/acsnano.4c14327},
	url = {https://doi.org/10.1021/acsnano.4c14327},
	volume = {19},
	year = {2025},
	year1 = {2025}}

@article{Fetida2024,
    Title = {Single-Spin Sensing: A Molecule-on-Tip Approach},
    Author = {Fétida, A. and Bengone, O. and Romeo, M. and Scheurer, F. and Robles, R. and Lorente, N. and Limot, L.}, 
    journal = {ACS Nano},
    year ={2024},
    pages = {13829- 13835}, 
    volume = {18},
    issue =  {21},
    Doi  = {doi: 10.1021/acsnano.4c02470},
    Url = {https://doi.org/10.1021/acsnano.4c02470}
    }

@article{Fetida2025,
    author = {Fétida, A. and Bengone, O. and Goyhenex, C. and Scheurer, F. and Robles, R. and Lorente, N. and Limot, L.},
    title = {Molecular spin-probe sensing of {H}-mediated changes in {C}o nanomagnets},
    journal = {Sci. Adv.},
    volume = {11},
    number = {7},
    pages = {eads1456},
    year = {2025},
    doi = {10.1126/sciadv.ads1456},
    URL = {https://www.science.org/doi/abs/10.1126/sciadv.ads1456}}

@article{Ormaza2017a,
    title = {Efficient spin-flip excitation of a Nickelocene molecule},
    journal = {Nano Lett.},
    volume = {17},
    issue ={3},
    pages = {1877-1882},
    year = {2017},
    doi = {10.1021/acs.nanolett.6b05204},
    url = {https://doi.org/10.1021/acs.nanolett.6b05204},
    author ={Ormaza, M. and Bachellier, N. and Faraggi, M. N. and Verlhac, B. and Abufager, P. and Ohresser, P. and Joly, L. and Romeo, M. and Scheurer, F. and Bocquet, M.-L. and Lorente, N. and Limot, L.}
}

@article{Ormaza2017b,
	Author = {Ormaza, M. and Abufager, P. and Verlhac, B. and Bachellier, N. and Bocquet, M. -L. and Lorente, N. and Limot, L.},
	Da = {2017/12/07},
	Doi = {10.1038/s41467-017-02151-6},
	Id = {Ormaza2017},
	Journal = {Nat. Commun.},
	Number = {1},
	Pages = {1974},
	Title = {Controlled spin switching in a metallocene molecular junction},
	Ty = {JOUR},
	Url = {https://doi.org/10.1038/s41467-017-02151-6},
	Volume = {8},
	Year = {2017}
	}

@article{Waeckerlin2022,
    author={W{\"a}ckerlin, C. and Cahl{\'i}k, A. and Goikoetxea, J. and Stetsovych, O. and Medvedeva, D. and Redondo, J. and {\v{S}}vec, M. and Delley, B. and Ondr{\'a}{\v{c}}ek, M. and Pinar, A. and Blanco-Rey, M. and Koloren{\v{c}}, J. and Arnau, A. and Jel{\'i}nek, P.},
    title={Role of the Magnetic Anisotropy in Atomic-Spin Sensing of 1{D} Molecular Chains},
    journal={ACS Nano},
    year={2022},
    month={Oct},
    day={25},
    publisher={American Chemical Society},
    volume={16},
    number={10},
    pages={16402-16413},
    issn={1936-0851},
    doi={10.1021/acsnano.2c05609},
    url={https://doi.org/10.1021/acsnano.2c05609}
}

@article{Song2024,
	author = {Song, S. and Pinar Sol{\'e}, A. and Mat{\v e}j, A. and Li, G. and Stetsovych, O. and Soler, D. and Yang, H. and Telychko, M. and Li, J. and Kumar, M. and Chen, Q. and Edalatmanesh, S. and Brabec, J. and Veis, L. and Wu, J. and Jelinek, P. and Lu, J.},
	date = {2024/06/01},
	doi = {10.1038/s41557-024-01453-9},
	id = {Song2024},
	isbn = {1755-4349},
	journal = {Nat. Chem.},
	number = {6},
	pages = {938--944},
	title = {Highly entangled polyradical nanographene with coexisting strong correlation and topological frustration},
	url = {https://doi.org/10.1038/s41557-024-01453-9},
	volume = {16},
	year = {2024}}

@article{Bae2025,
	author = {Bae, Y. and Ternes, M. and Yang, K. and Heinrich, . J. and Wolf, C. and Lutz, C. P.},
	date = {2025/01/14},
	doi = {10.1021/acsnano.4c13934},
	isbn = {1936-0851},
	journal = {ACS Nano},
	journal1 = {ACS Nano},
	journal2 = {ACS Nano},
	month = {01},
	number = {1},
	pages = {1361--1370},
	publisher = {American Chemical Society},
	title = {Direct Observation of Fully Spin-Polarized Tunnel Current Between Quantum Spins Using a Single Molecule Sensor},
	type = {doi: 10.1021/acsnano.4c13934},
	url = {https://doi.org/10.1021/acsnano.4c13934},
	volume = {19},
	year = {2025},
	year1 = {2025}}

@article{Lorenz2026,
    author = {Meyer, L. and Kögler, M. and Henninger, R. and Néel, N. and Kröger, J.},
    title = {Crossing and Anticrossing of Exchange-Coupled Molecular Spin Excitation Energy Levels},
    journal = {Small},
    volume = {22},
    number = {12},
    pages = {2412703},
    doi = {https://doi.org/10.1002/smll.202412703},
    url = {https://onlinelibrary.wiley.com/doi/abs/10.1002/smll.202412703},
    year = {2026}
}

@article{Bachellier2016,
    title = {Unveiling nickelocene bonding to a noble metal surface},
    author = {Bachellier, N. and Ormaza, M. and Faraggi, M. and Verlhac, B. and V\'erot, M. and Le Bahers, T. and Bocquet, M.-L. and Limot, L.},
    journal = {Phys. Rev. B},
    volume = {93},
    issue = {19},
    pages = {195403},
    numpages = {6},
    year = {2016},
    month = {5},
    publisher = {American Physical Society},
    doi = {10.1103/PhysRevB.93.195403},
    url = {https://link.aps.org/doi/10.1103/PhysRevB.93.195403}
}

@article{Chen1990,
    title = {Tunneling matrix elements in three-dimensional space: The derivative rule and the sum rule},
    author = {Chen, C. Julian},
    journal = {Phys. Rev. B},
    volume = {42},
    issue = {14},
    pages = {8841--8857},
    numpages = {0},
    year = {1990},
    month = {Nov},
    publisher = {American Physical Society},
    doi = {10.1103/PhysRevB.42.8841},
    url = {https://link.aps.org/doi/10.1103/PhysRevB.42.8841}
}

@article{Ternes2015,
  	author={M. Ternes},
  	title={Spin excitations and correlations in scanning tunneling spectroscopy},
  	journal={New J. Phys.},
  	volume={17},
  	number={6},
  	pages={063016},
  	url={http://stacks.iop.org/1367-2630/17/i=6/a=063016},
  	year={2015}
  	}

@article{Loth2010b,
    doi = {10.1088/1367-2630/12/12/125021},
    url = {https://dx.doi.org/10.1088/1367-2630/12/12/125021},
    year = {2010},
    month = {dec},
    publisher = {},
    volume = {12},
    number = {12},
    pages = {125021},
    author = {S. Loth and C. P. Lutz and A. J. Heinrich},
    title = {Spin-polarized spin excitation spectroscopy},
    journal = {New J. Phys.}
}

@article{Atodirisei2010,
    title = {Design of the Local Spin Polarization at the Organic-Ferromagnetic Interface},
    author = {Atodiresei, N. and Brede, J. and Lazi\'{c}, P. and Caciuc, V. and Hoffmann, G. and Wiesendanger, R. and Bl\"ugel, S.},
    journal = {Phys. Rev. Lett.},
    volume = {105},
    issue = {6},
    pages = {066601},
    numpages = {4},
    year = {2010},
    month = {Aug},
    publisher = {American Physical Society},
    doi = {10.1103/PhysRevLett.105.066601},
    url = {https://link.aps.org/doi/10.1103/PhysRevLett.105.066601}
}

@article{Kogler2024,
	author = {K{\"o}gler, M. and N{\'e}el, N. and Limot, L. and Kr{\"o}ger, J.},
	date = {2024/11/13},
	doi = {10.1021/acs.nanolett.4c04075},
	isbn = {1530-6984},
	journal = {Nano Lett.},
	month = {11},
	number = {45},
	pages = {14355--14362},
	publisher = {American Chemical Society},
	title = {Structural Manipulation of Spin Excitations in a Molecular Junction},
	type = {doi: 10.1021/acs.nanolett.4c04075},
	url = {https://doi.org/10.1021/acs.nanolett.4c04075},
	volume = {24},
	year = {2024},
	year1 = {2024}}

@Article{Bardeen1961,
  author    = {J. Bardeen},
  journal   = {Phys. Rev. Lett.},
  title     = {Tunnelling from a Many-Particle Point of View},
  year      = {1961},
  month     = {jan},
  number    = {2},
  pages     = {57--59},
  volume    = {6},
  doi       = {10.1103/physrevlett.6.57},
  publisher = {American Physical Society ({APS})},
}

 \end{document}